\documentclass[a4paper]{article}
\usepackage{amsmath,amssymb,amsfonts}
\usepackage{algorithmic}
\usepackage[pdftex]{graphicx}
\usepackage{textcomp}
\usepackage{xcolor}
\usepackage{url}

\usepackage{INTERSPEECH2019}

\title{Audio Classification of Bit-Representation Waveform}
\name{Masaki Okawa$^1$, Takuya Saito$^1$, Naoki Sawada$^{1,2}$, Hiromitsu Nishizaki$^1$}
\address{
  $^1$Integrated Graduate School of Medicine Engineering, and Agricultural Sciences, \\ University of Yamanashi, \\
  $^2$Idein Inc.}
\email{\{yukari\_phantasm, saitoh\_t, sawada\}@alps-lab.org, hnishi@yamanashi.ac.jp}

\begin{document}

\renewcommand{\baselinestretch}{0.92}
\textfloatsep 6pt
\floatsep 5pt

\maketitle
\begin{abstract}
This study investigated the waveform representation for audio signal classification. Recently, many studies on audio waveform classification such as acoustic event detection and music genre classification have been published. Most studies on audio waveform classification have proposed the use of a deep learning (neural network) framework. Generally, a frequency analysis method such as Fourier transform is applied to extract the frequency or spectral information from the input audio waveform before inputting the raw audio waveform into the neural network. In contrast to these previous studies, in this paper, we propose a novel waveform representation method, in which audio waveforms are represented as a bit sequence, for audio classification. In our experiment, we compare the proposed bit representation waveform, which is directly given to a neural network, to other representations of audio waveforms such as a raw audio waveform and a power spectrum with two classification tasks: one is an acoustic event classification task and the other is a sound/music classification task. The experimental results showed that the bit representation waveform achieved the best classification performance for both the tasks.
\end{abstract}

\section{Introduction}

Recently, there have been many studies \cite{DCASE2018, IEEESP, GTZAN01} on audio/music detection or classification tasks using deep neural networks (DNNs)\cite{DCASE01,DCASE02,DCASE03,ETRI,IEEEACM,CRNN} because environmental sound analysis including detection and classification will become a key technology in the near future.

In acoustic event detection tasks, some corpora \cite{TUT, ESC, sony, audioset} for event detection have been released and they are used at competitions, such as DCASE\footnote{\url{http://dcase.community}}. Additionally, for the music genre classification task, GTZAN \cite{GTZAN01} and FMA \cite{FMA} have been released. Thus, we can use many datasets for studying audio/sound/music processing.

Generally, in an audio detection task, a raw audio waveform is pre-processed before inputting it into a neural network. Many pre-processing methods for a raw audio waveform have been tried to use. For example, Mel-Frequency Cepstrum Coefficients (MFCCs) and filter bank outputs\cite{FB} are popular and widely used. Most pre-processing approaches transform a time domain waveform into its frequency domain by computing the Discrete Fourier Transform (DFT). Recently, end-to-end approaches have become popular; however, there are few studies \cite{Lee01, SCNN, IS2016} that have investigated feature extraction methods for a time-domain waveform. This is because a neural network architecture for audio classification has mainly two hierarchies: one is composed of convolutional neural network (CNN) layers for feature extraction and the other consists of recurrent-based layer(s) and/or fully-connected (FC) layer(s) to classify the input waveform. The CNN layer(s) can extract features from the input waveform instead of human-designed features. Until recently, therefore, some studies tackled the pre-processing methods of an audio waveform. Lee et al. \cite{Lee01, SCNN} proposed a CNN architecture that can learn representations using sample-level filters on a music genre classification task. Sainath et al. \cite{IS2016} also investigated raw waveforms that were directly inputted into the convolution layer.

In contract to the previous works, we propose a new pre-processing method that converts a raw audio waveform into a bit-based representation. We investigated two sorts of bit representations of an audio waveform in this study. This is a novel idea that has never proposed before. We evaluated the effectiveness of the bit representation of the raw audio by directly classifying the audio files. 

The contributions of this paper are as follows:
\vspace*{0mm}
\begin{itemize}
  \itemsep -2pt
\item The paper first shows that a bit representation waveform forms convolution layers of a neural network to extract more effective features from the audio. Therefore, the performances of both the acoustic event detection task and the music/speech binary classification task considerably improved compared to typical data representations, such as MFCCs and a power spectrum. 
\item The study experimentally shows that a bit representation waveform has noise robustness for a music/sound classification task because the time sequences of higher-order bits of the bit representation waveform were not strongly affected by noise.
\item The bit representation also has robustness against the domain mismatch between training and testing of the neural network.
\item The bit representation of a raw audio waveform is useful for various types of audio classification. 
\end{itemize}

This paper is organized as follows: the next section presents how to transform an audio waveform into a bit representation waveform. Section 3 describes the neural network architectures used for two types of audio classification tasks. Section 4 describes the experimental setups for the two tasks and the results, and our conclusions are discussed in Section 5.

\section{Bit Representation of a Raw Audio Waveform}

In this paper, we introduce the use of bit representations of a raw audio waveform. Two types of bit representations of a raw audio waveform will be reviewed. 

Generally, a sampled value is represented as a signed integer. For example, the sampled value varies from $-32768$ to $+32767$ when the quantization bit rate is 16 bits per sample. A previous study \cite{SCNN}, which dealt with a raw sound waveform on the music genre classification task, used a raw-sampled value sequence as the input representation to the neural network. Unlike that previous study \cite{SCNN}, our method transforms a sampled integer value into a bit vector, which is a set of bit values of the sampled value. For example, when the sampled integer is ``$12$'' (quantization bit rate is 8/sample) , the bit vector of ``$12$'' is shown as ``$(0,0,0,0,1,0,1,0)$.'' The sequence of bit vector sequence proceeds into a neural network.

The bit representation transformation process can make a more rich representation of a raw audio waveform compared to an integer-based audio waveform representation. Therefore, we conjecture that convolution layers in a classifier can extract more effective feature maps from a bit representation waveform to achieve highly accurate audio detection/classification. 

\subsection{Bit pulse extraction}

\begin{figure}[t]
\centering
\includegraphics[width=0.95\columnwidth]{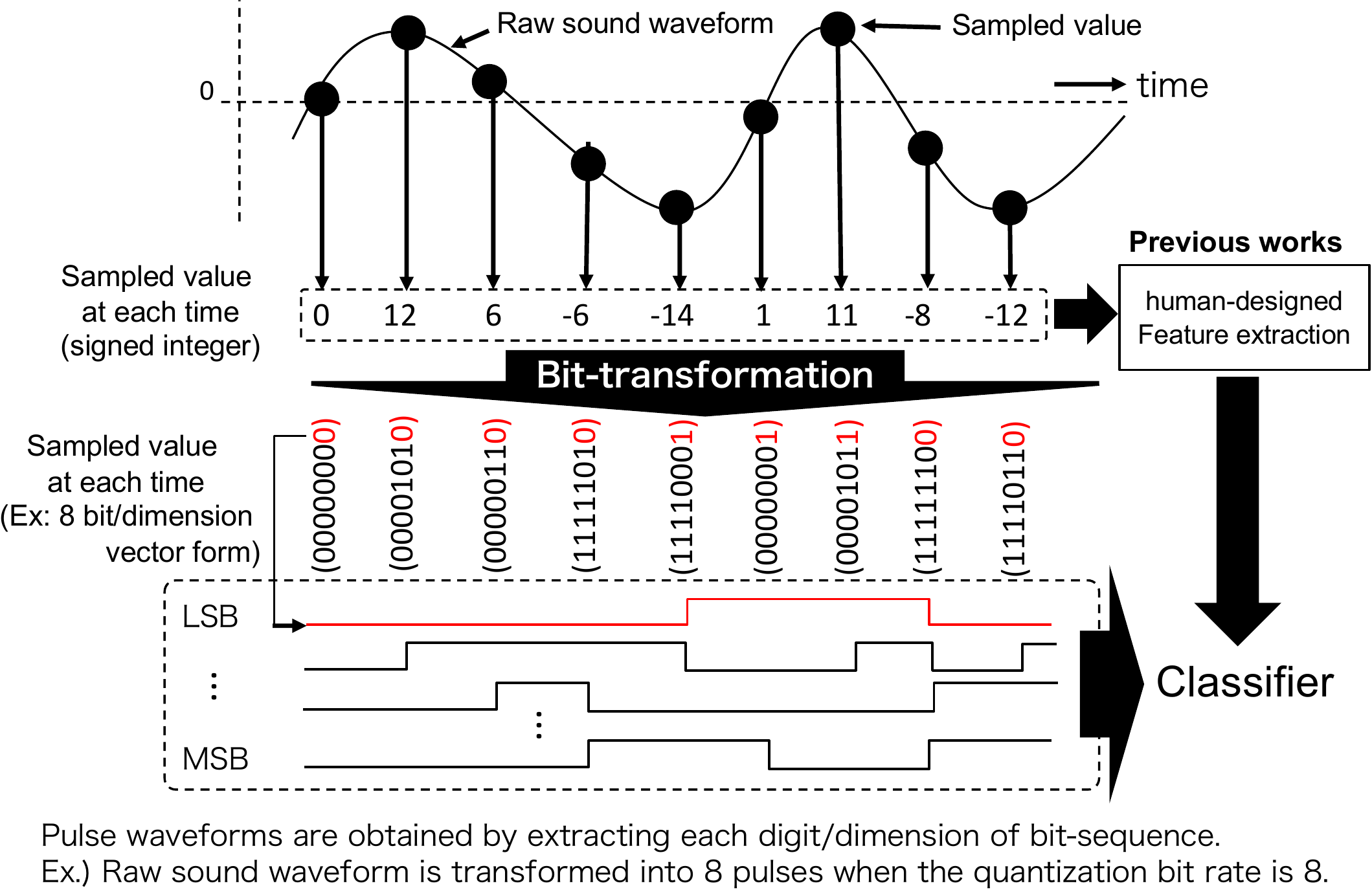}
\caption{Extraction of bit-based pulse waveforms from a raw sound waveform. }
\label{fig:bit01}
\end{figure}

Figure \ref{fig:bit01} shows one of the methods for bit transformation from a raw audio waveform. Figure \ref{fig:bit01} shows an example when the quantization bit rate is 8/sample. In this transformation method, each sampled integer value at each time $i$ (the total number of samples is $T$) is expanded to a bit representation vector; then, bit pulse waveforms are obtained by extracting each dimension/digit of the bit representation vector sequence.
Finally, the bit-pulse waveforms are inputted to a CNN layer of a classifier, where ONE waveform is treated as ONE input channel.

The main idea of this method is to segment a raw audio waveform (integer value sequence) into several pulse waveforms. For example, as shown in Fig. \ref{fig:bit01}, we obtaine eight bit pulse waveforms when the quantization bit rate is 8/sample. The bit pulse waveform that is based on the Most Significant Bit (MSB) can capture the dynamic change (plus or minus) of the raw audio waveform. Like the MSB example, each bit pulse waveform has a role for extracting audio features. Therefore, we can accurately analyze the raw audio waveform using bit representation transformation in constract to when the transformation process is not used.

\subsection{Bit pattern image}

\begin{figure}[t]
\centering
\includegraphics[width=0.95\columnwidth]{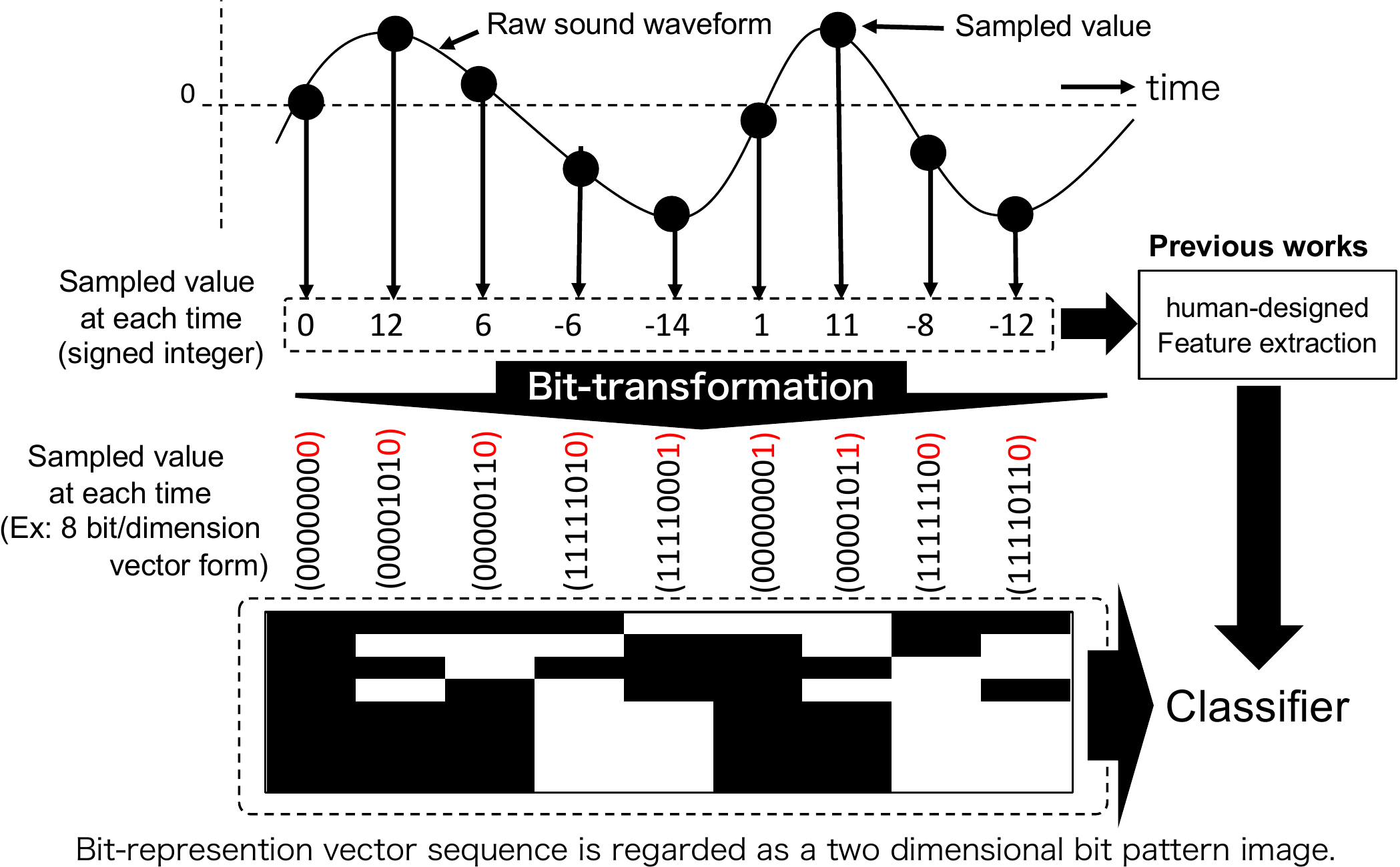}
\caption{Extraction of the bit pattern image from a raw audio waveform.}
\label{fig:bit02}
\end{figure}

The other method for transforming a raw audio waveform into a bit representation-based sequence is to convert a waveform into a two-dimensional bit pattern image. Figure \ref{fig:bit02} also shows the other bit representation method for a raw audio waveform. This figure is the same as Fig. \ref{fig:bit01} regarding the quantization bit rate.  Unlike the transformation method shown in Fig. \ref{fig:bit01}, the method shown in Fig. \ref{fig:bit02} creates a two-dimensional bit pattern image from the bit representation vector sequence. A bit pattern image is given to the two-dimensional convolution layer.

Some believe that the image representation from a raw audio waveform is used for sound classification tasks \cite{ASC,IS2014} or for speech recognition (key word spotting) tasks \cite{XXX}. These studies used a sound spectrogram image from the audio waveform. Our proposed method is entirely different from previous methods \cite{ASC,IS2014,XXX}. Our method does not lose the information included in the raw audio waveform. 

\section{Neural Networks}

In this study, we use two sorts of neural networks depending on the evaluation tasks, and the details are provided in Section 4. This section explains these neural network architectures. There are infinite neural network architectures. Of note, the neural network architectures adopted in this paper are not necessarily optimal for these tasks. We should explore an optimal neural network architecture is future work. The main purpose of this study is to investigate the data representation for neural network-based classifiers.

\subsection{Neural network for a bit pulse waveform} \label{sec:nn01}

\begin{figure}[tb]
\centering
\includegraphics[width=0.9\columnwidth]{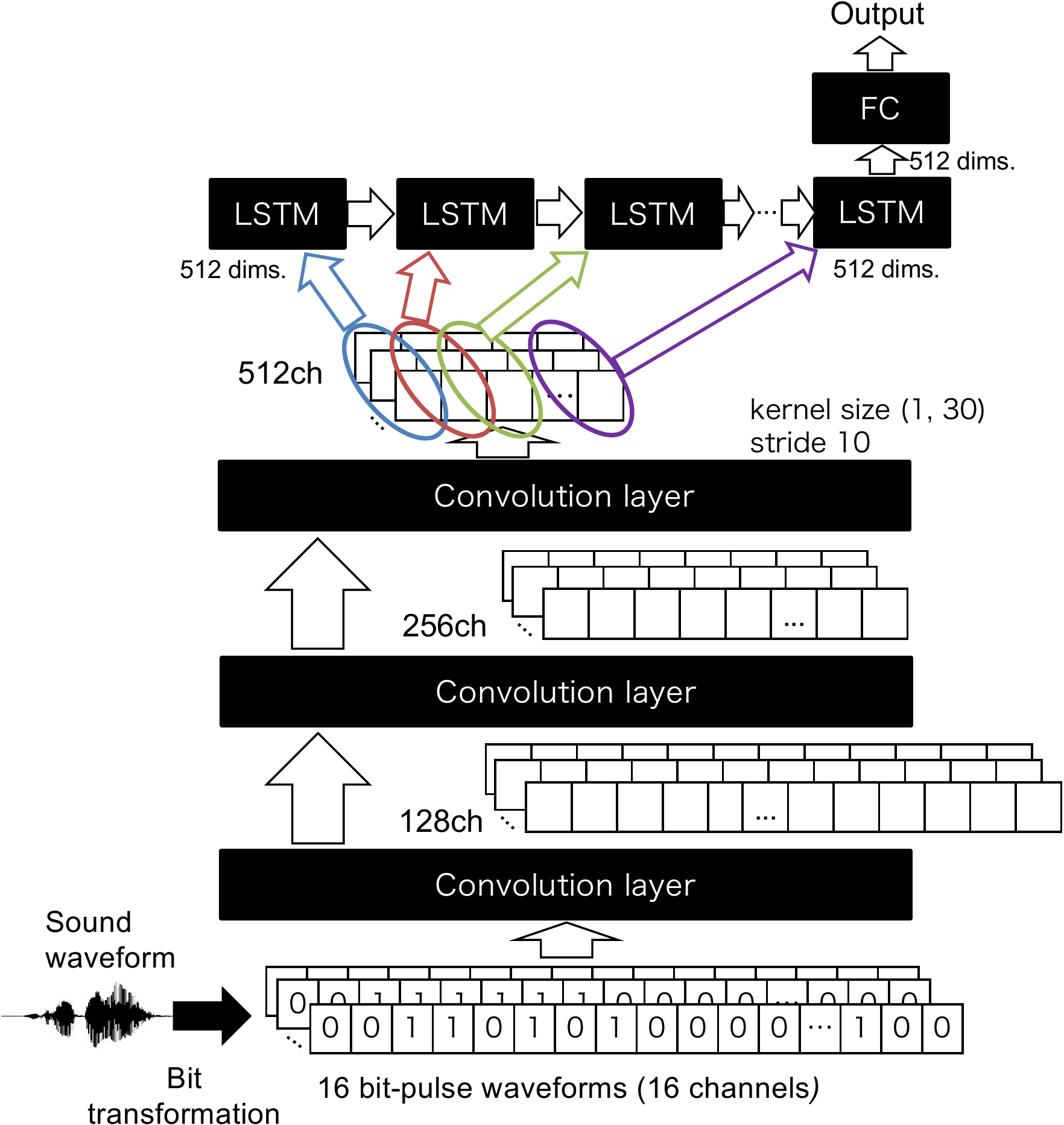}
\vspace*{-2mm}
\caption{Neural network architechture for the bit pulse waveform of the audio waveform.}
\label{fig:nn01}
\end{figure}

Figure \ref{fig:nn01} shows a CNN Long Short-Term Memory (LSTM) network architecture that can accept a bit representation waveform that has multi-channels. This is a two-stage model that is widely used for classification or recognition of time sequential data. The CNN layers extract a feature map from the input waveforms, and the LSTM layer with the FC layer classifies the input waveform using historical information.

Bit pulse waveforms (multi-channel waveforms) are inputted to the first CNN layer, and then the third CNN layer outputs the feature map of the 512 channel with size of (1, $N$). The number of $N$ specifically depends on the length (duration) of the input waveform. The feature map is divided into $N$ small feature maps each of which has 512 dimensions. Each small feature map proceeds to the LSTM layer in sequential order. Finally, the output of the LSTM layer, when the final ($N$-th) small feature map is given to the LSTM, is transferred to the FC layer.



\subsection{Neural network for the bit pattern image}

\begin{figure}[tb]
\centering
\includegraphics[width=0.85\columnwidth]{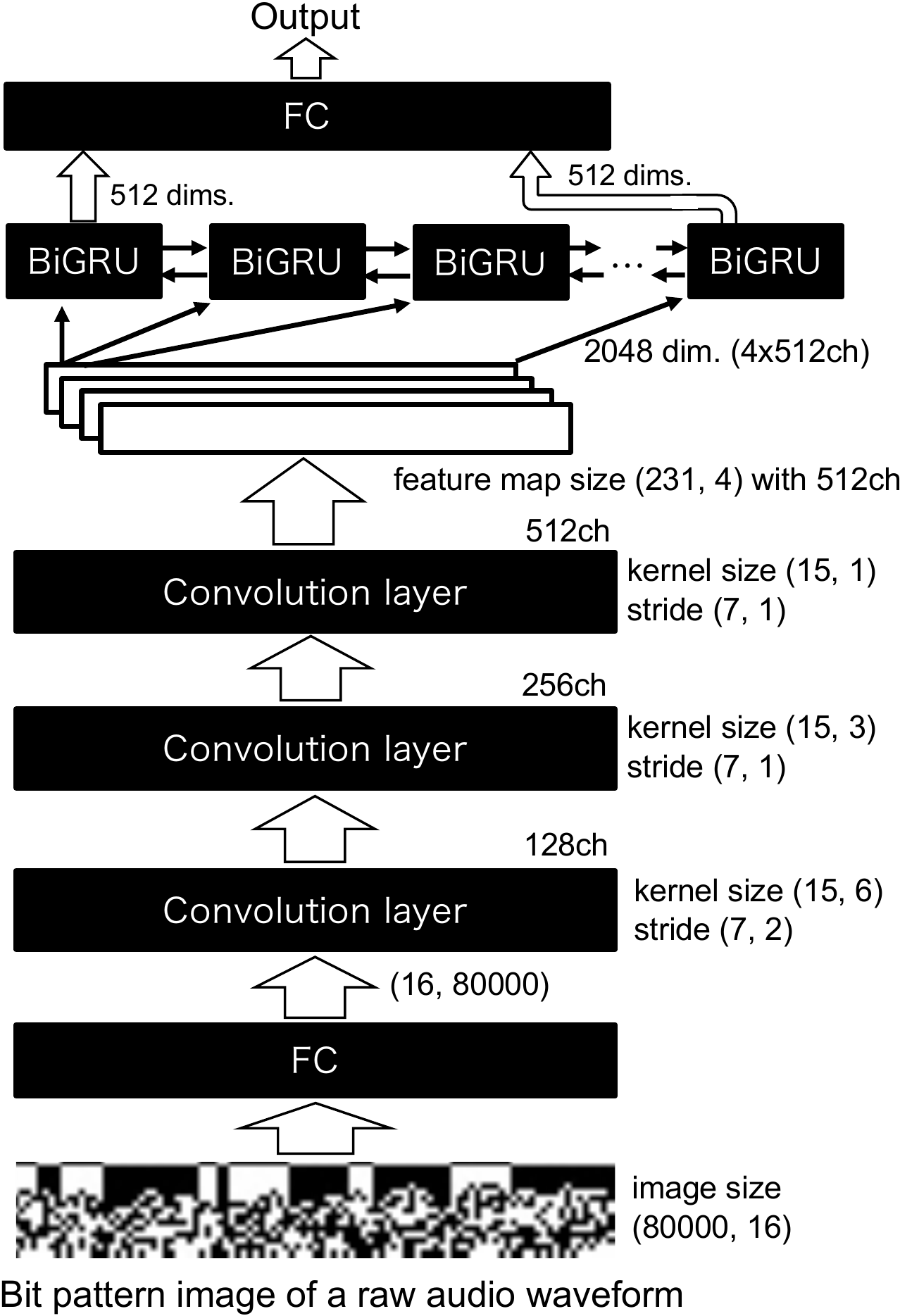}
\vspace*{-2mm}
\caption{Neural network architechture for the bit pattern image of the raw audio waveform.}
\label{fig:nn02}
\end{figure}

Figure \ref{fig:nn02} also displays a CNN Bi-directional Gated Recurrent Unit (BiGRU) network architecture that can accept the bit pattern image of an audio waveform. The model architecture is a little similar than the CNN-LSTM network described in Section \ref{sec:nn01}; however, this architecture has a BiGRU layer in constrast to the CNN-LSTM. In addition, each CNN layer performs a two-dimensional convolution operation. The role of the lower FC layer is to change the bit scale.

The CNN layers finally extract a feature map that has a size of (231, 4) with 512 channels from the bit pattern image. This feature map is divided into 231 small feature maps, each of which has 2,048 ($=$4$\times$512) dimensions. Each small feature maps is inputted into the BiGRU layer in sequential and reverse sequential orders.

\section{Experiments}

Our bit transformation method and the neural network architecture were evaluated using two tasks: one was an acoustic event detection task and the other was a music/speech classification task.

\subsection{Experimental setup}
\subsubsection{Acoustic event detection task}

We used the acoustic event classification database\footnote{\url{https://data.vision.ee.ethz.ch/cvl/ae dataset}} harvested from Freesound\footnote{\url{http://www.freesound.org/}} for the acoustic event detection task in this study. The database consists of 28 events. The details of the dataset are described in Takahashi's paper \cite{sony}.

The total number of audio files was 5,223, and its duration was 768.4 min. The dataset was separated into two groups: one was for training the CNN-LSTM model and the other was for testing. It included 3,702 training and 1,521 testing audio files. All audio files were converted to a 16,000Hz sampling rate with a 16-bit quantization rate. The condition for the data separation was exactly the same as in the previous study \cite{sony}. Therefore, we could compare the results between our methods and those of Takahashi's result \cite{sony}. 

The CNN-LSTM model used for this task was for the bit pulse-based transformation, as shown in Fig. \ref{fig:nn01}. Therefore, an input audio waveform was separated into 16 bits-based pulse waveforms (Fig. \ref{fig:bit01}) before the audio proceeded to the neural network. The training condition of the CNN-LSTM is described in Table \ref{tbl:traincond01}. The hyper-parameters such as kernel size in each CNN layer are heuristically decided.

\begin{table}[tb]
\centering
\caption{Training condition of the CNN-LSTM model for pulse waveform.}
\label{tbl:traincond01}
\vspace*{-2mm}
\begin{tabular}{l|l} \hline\hline
Mini-batch size & 64 \\
Num. of  epochs & 500 \\
Kernel size for the CNN layers & (1, 30) \\
Stride for the CNN layers & (1, 10) \\
Channels (1st, 2nd, 3rd CNN) & (128, 256, 512) \\ 
Activation for the CNN layers & ReLU \cite{ReLU} \\
Dropout & 0.2 at all the CNN layers \\
Batch normalization\cite{Ioffe/2015/arXiv} & No \\
Loss func. & Softmax cross entropy \\
Optimizer & MomentumSGD \\
Learning rate & 0.002 \\ \hline\hline
\end{tabular}
\end{table}

For this task, we compared our bit representation method to the typical data representations: MFCC, power spectrum, and raw sound waveform (integer representation). The neural network model with the power spectrum representation is the same as that in ref. \cite{sony}. Of note, this model is different from Fig. \ref{fig:nn01}. The 256-dimensional power spectrum and its $\Delta$ and $\Delta\Delta$ were inputted to the CNN layer. In the case of MFCC and the raw sound waveform, the network architechtures were the same as in Fig. \ref{fig:nn01}; however, the parameters were optimized for each input feature.

\subsubsection{Music/speech classification task}

\begin{table}[tb]
\centering
\caption{Training condition for the CNN-BiGRU model for the bit pattern image.}
\label{tbl:traincond02}
\vspace*{-2mm}
\scalebox{0.87}{
\begin{tabular}{l|l} \hline\hline
Mini-batch size & 32 \\
Num. of  epochs & 300 \\
Kernel size for the CNN layers & See Fig.\ref{fig:nn02} \\
Stride for the CNN layers & See Fig.\ref{fig:nn02} \\
Channels (1st, 2nd, 3rd CNN) & (128, 256, 512) \\ 
Activation for the FC and CNN layers & ReLU\cite{ReLU} \\
Dropout & 0.2 for all the CNN layers \\
Batch normalization \cite{Ioffe/2015/arXiv} & No \\
Loss func. & Softmax cross entropy \\
Optimizer & MomentumSGD \\
Learning rate & 0.01 (half every 30 epochs) \\ \hline\hline
\end{tabular}
}
\end{table}

For the second music/speech binary classification task, we used the Marsyas dataset\cite{GTZAN}. This was also used in a previous study \cite{2014/EUSIPCO}. The dataset consists of music and speech classes each of which has 64 audio files. Each audio file was in a WAVE format and the sampling frequency rate was 22,050Hz with a 16-bit quantization rate. Each audio file was segmented into 10 seconds audio files, and they were converted to a 8,000Hz sampling frequency with a 16-bit quantization rate. Finally, the total number of audio files was 384, and they were separated into two groups: one group (269 files) was for training the CNN-BiGRU and the other (115) was for testing. The 10 seconds waveform was converted to an (80000, 16)-sized bit pattern image.

The neural network model used for the second task was for the bit pattern image, as shown in Fig. \ref{fig:nn02}. Therefore, an input audio waveform (10 seconds file) was converted to a (80000, 16)-sized bit pattern image before the audio file proceeded to the neural network. The training condition of the CNN-BiGRU is described in Table \ref{tbl:traincond02}.

In this task, we investigate the domain dependency of the CNN-BiGRU model that accepts the bit pattern image of the audio waveform. When we used raw information from the sound waveform, we were concerned that the neural network may have been over trained. 
Therefore, we collected audio files from an radio broadcast in Japan; the total number of radio files was 400 from a music/speech label. The duration of each audio file was 10 seconds which was the same as in the Marsyas dataset. Then, another classification model, whose architecture was the same as that of the CNN-BiGRU, was trained using the radio audio files. Because this model was trained from the out-of-domain dataset for testing, we could investigate the domain dependency of the model using the bit pattern image transformation method.

Moreover, we investigated the noise robustness of the bit representation. We summarize the experimental conditions as follows:
\begin{description}
\itemsep 0pt
\item[C01]: The model was trained from the Marsyas dataset, and it was evaluated using 115 files (clean).
\item[C02]: The same model was used as for C01; however, it was evaluated using 115 files to which 10 dB white noise was injected. 
\item[C03]: The model was trained from the radio files, and it was evaluated using 115 files (clean).
\end{description}


\subsection{Results and discussion}

Table \ref{tbl:result01} shows the classification accuracy rates for each data representation for the acoustic event detection task. We compared four sorts of data representations. As shown in Table \ref{tbl:result01}, the bit representation of the audio waveform exhibited the best accuracy rate of 88.4\% compared to the other (typical) data representation methods. We deduced that the CNN layers can extract more effective feature maps from bit pulse waveforms because bit representation allows the CNN layers to detect a minute change in the audio waveform more easily than a typical data representation, such as power spectrum.

\begin{table}[tb]
\begin{center}
\caption{Classification accuracy rates for each data representation for the acoustic event detection task.}
 \label{tbl:result01}
 \vspace*{-2mm} 
\begin{tabular}{l||c}
\hline
Data representation & Accuracy [\%] \\
\hline\hline
Bit-representation (bit-pulse) & \underline{\bf 88.4} \\
MFCC & 57.4 \\
Power spectrum \cite{sony} & 80.3$^*$ \\
Raw waveform (integer) & 34.6 \\ \hline
\end{tabular} \\
$^*$ This value is from Table 3 in ref. \cite{sony}.
\end{center}
\end{table}

For the second task, which was the music/speech binary classification task, the classification accuracy rates were determined and are summarized in Table \ref{tbl:result02}. Under the {\bf C01} condition (matched condition), the raw audio waveform was slightly better than the bit representation. However, there were no significant differences between them. Additionally, raw-level representations exhibited much better performances against the power spectrum and MFCC\footnote{However, the model architecture for MFCC is different from the one used in this study. Therefore, we could not compare them.}.

Under the {\bf C02} condition, the bit representation was considerably better than the raw audio waveform. The raw waveform was subject to effects of noise; however, the bit representation was more robust in the noisy environment. Finally, for the mismatched condition between training and testing ({\bf C03}), the bit representation also exhibited robustness because the accuracy rate was slightly damaged.

\begin{table}[tb]
\begin{center}
 \caption{Segment-based classification accuracy rates [\%] of each data representation for the music/speech classification task.}
 \label{tbl:result02}  
\vspace*{-2mm} 
\begin{tabular}{l||c|c|c}
\hline
Data representation & \multicolumn{3}{c}{Experimental conditions} \\ \cline{2-4}
                    & C01 & C02 & C03 \\ \hline\hline
Bit representation & 94.7 & \underline{\bf 81.2} & \underline{\bf 92.8} \\
Raw waveform (integer) & \underline{95.4} & 71.8 & 83.0 \\
Power spectrum & 90.5 & --- & --- \\
MFCC with RBM \cite{2014/EUSIPCO} & 91$^*$ & --- & --- \\
\hline
\end{tabular} \\
$^*$ This value is from Fig.2 (b) in ref. \cite{2014/EUSIPCO}.
\end{center}
\end{table}


As shown in Table \ref{tbl:result01} and \ref{tbl:result02}, the bit representation for audio waveform is very useful for the audio classification tasks from the viewpoint of feature map extraction and robustness for noise and training conditions of the neural network. We also applied the bit representation to a MNIST (hand-writing number classification) task \cite{MNIST} using the VGG-16 network \cite{VGG16} as an additional experiment to investigate the effectiveness of the bit-representation of the data. The bit representations for the hand-writing images achieved a 99.73\% classification accuracy compared to 99.58\% from for gray-scale images. The bit representation may be adequate for any image classification task.

\vspace*{-3mm}
\section{Conclusions}

This paper proposed a novel data representation method for audio classification using a deep neural network. The proposed method transforms a raw audio waveform (integer value sequence) into either bit-based multi-channel pulse waveforms or a bit pattern image depending on the classification task before inputting to a neural network. The experimental results of both the tasks showed that the bit representation of an audio waveform achieved the best results compared to the other data preprocessing methods.

In future work, we will explore the optimal neural network architecture for a bit representation waveform for various audio sets. Moreover, we will also apply our method to other tasks such as speech recognition and speaker recognition.

\section{Acknowledgments}
This work was supported by JSPS KAKENHI Grant-in-Aid for Scientific Research (B) Grant Number 17H01977.



\end{document}